\documentstyle[prl,aps,twocolumn,epsf]{revtex}

\def\epsfig#1#2#3#4
         {
         \epsfysize=#2 \vbox{ \hglue#3 \epsfbox[#4]{#1} }
         }

\begin{document}
\draft
\title{Dynamical simulation of transport in one-dimensional quantum wires}
\author{Kevin Leung,$^1$ Reinhold Egger,$^2$ and C.H. Mak$^1$}
\address{${}^1$Department of Chemistry, University of Southern California,
Los Angeles, CA 90089-0482, USA\\
${}^2$Fakult\"at Physik, Albert-Ludwigs-Universit\"at, 
Hermann-Herder-Stra{\ss}e 3, D-79104 Freiburg, Germany} 
\date{Date: \today}
\maketitle

\begin{abstract}
Transport of single-channel spinless interacting fermions 
(Luttinger liquid) through a barrier has been studied 
by numerically exact quantum Monte Carlo methods. 
A novel stochastic integration over the real-time paths
allows for direct computation of nonequilibrium conductance and 
noise properties.  We have examined the low-temperature
scaling of the conductance in the crossover region
between a very weak and an almost insulating barrier.
\end{abstract}


\narrowtext

Studies of transport in one-dimensional (1D) 
interacting Fermi systems have gathered
novel physical insights and led to the development of several new techniques
over the past few years.  Instead of the usual Fermi liquid theory,
the low-temperature properties of an idealized 1D quantum wire
are described by the Luttinger liquid model\cite{lutt64,hal81}.
Remarkable transport behaviors are expected for a Luttinger 
liquid in the presence of impurities or barriers,
and several interesting theoretical \cite{kf92,glaz,fls95,moon93,egger95} 
and experimental \cite{ths95,mill} studies have emerged recently. 
In this Letter, we describe a real-time quantum Monte Carlo (QMC) method 
that enables us to directly address the dynamics of a Luttinger liquid,
and use it to investigate the low-temperature transport properties of 
a 1D quantum wire.  

So far the only numerical approach to this problem has been the
one given by
Moon {\em et al.}\cite{moon93}. These workers examined 
transport in a Luttinger liquid using Euclidean-time QMC simulations by
analytically continuing numerical data to real time with Pad\'e 
approximants. Mathematically, this analytic continuation is ill-posed and
small statistical errors in the imaginary-time data
could be magnified immensely\cite{gube}.
Physically, imaginary-time data collected at low temperatures contain 
predominantly ground-state information whereas real-time dynamics
is controlled by the excitation spectrum.
Although the findings of Ref.\cite{moon93} are certainly reasonable,
a direct real-time simulation which avoids the
possibly troublesome analytic continuation is desirable.
Furthermore, a real-time simulation 
needs not rely on the Kubo formula and 
allows for direct computations of nonequilibrium
transport and noise properties beyond the reach of previous methods.

Common to all real-time QMC simulations is the ubiquitous and fundamental 
{\em dynamical sign problem}\cite{gube,filinov,doll1,makri,makprl,pre94}. 
It arises
in the stochastic summation over the system paths
when the real-time propagators are oscillatory. The
quantum-mechanical interference between different paths leads to a 
vanishing signal-to-noise ratio at very long real times and the 
simulation becomes effectively unstable. However, it is generally possible to 
treat intermediate-to-long times by employing a {\em partial summation
scheme}\cite{makprl} which reduces the effective number of variables 
subject to the Monte Carlo sampling.  If guided by physical intuition, such a 
partial summation scheme can largely circumvent the dynamical sign problem.
Previous applications to dissipative tight-binding models
in the quantum-chemical context \cite{makprl,pre94}
have demonstrated the practical usefulness of this concept.
Here, we adapt this idea to transport in interacting 1D quantum wires,
and as we demonstrate below, the method is able to extract 
numerically exact results in the interesting low-temperature region.

Transport in Luttinger liquids depends crucially
on a dimensionless parameter $g$ which characterizes
the electron-electron interaction strength. For the 
non-interacting case, $g=1$.  For $g<1$, interactions are repulsive,
and a barrier of any height leads to a perfect insulator at zero
temperature.  At asymptotically low temperatures, Kane and 
Fisher (KF) \cite{kf92} have predicted the following scaling
 behavior for the linear dc conductance: 
\begin{equation}
G \sim T^{2/g-2} \;.	\label{kfscale}
\end{equation}

Most analytical methods applied so far assume either a very small
or a very large barrier height (impurity level) $V_0$ 
compared to the fermion bandwidth $\omega_{\rm c}$.
For instance, perturbative arguments lead to Eq.(\ref{kfscale})
in the almost insulating case $V_0/\omega_{\rm c} \gg 1$, while 
exact results \cite{kf92,glaz,fls95,egger95} find the same scaling law
for the opposite limit $V_0/\omega_{\rm c}\ll 1$.
Unfortunately, little is known in the crossover region between the
two.  A recent study\cite{eg95}
has shown that the usual bosonized impurity Hamiltonian 
\cite{kf92} can be used to describe this crossover
and is valid even for
$\pi V_0/\omega_{\rm c} > 1$, which 
corresponds to an impurity level outside the fermion band.
Interestingly, Guinea {\em et al.}\cite{guin95} have suggested
a different low-temperature scaling form for the weak-barrier case
\begin{equation}
G = c_1 T^2 + c_2 T^{2/g-2} \label{crossover}\;.
\end{equation}
While exact results\cite{kf92,glaz,fls95,egger95} give
$c_1=0$ for both $V_0/\omega_{\rm c} \gg 1$ and $V_0/\omega_{\rm c}\ll 1$,
they cannot rule out the $T^2$ scaling term for {\em intermediate} barrier
heights $\pi V_0/\omega_{\rm c}\approx 1$.
A naive reasoning leading to Eq.(\ref{crossover}) for intermediate
barriers is as follows.
While the anomalous KF scaling exponents hold near
the quantum critical points $V_0 /\omega_{\rm c}=0$ and $\infty$,
scale invariance may not survive far away.
One might then expect regular analytical behavior $G \sim T^n$
with even power $n$, and the lowest-order term would give $T^2$
as shown in (\ref{crossover}).
If this scaling form is indeed true, at sufficiently 
low temperatures the $T^2$ scaling
would dominate for $g<1/2$ and the
KF law (\ref{kfscale}) could only be observed for $g>1/2$.  
This prediction is clearly of practical importance
for future experiments on quantum wires. As an application of our 
QMC method, we will address the validity of (\ref{crossover}) here.

The action for the Luttinger liquid transport problem is usually 
formulated in terms of a phase field $\theta$ \cite{hal81}.
In the presence of a short-ranged 
impurity potential of effective strength (barrier height) $V_0$, the 
(real-time) Lagrangian density takes the form \cite{kf92}
\begin{eqnarray} \label{lag1}
{\cal L}&&=
(1/2g)\,\partial_\mu \theta(x,t)\,\partial^\mu \theta(x,t)\\
 +&& V_0\, \delta(x) \cos[2\sqrt{\pi} \theta(x,t)]  + (eV /\sqrt{\pi})\,
\delta(x) \,\theta(x,t) \;, \nonumber
\end{eqnarray}
where $V$ denotes an external static voltage. 
Since one can eliminate all degrees of freedom away from the barrier
by Gaussian integrations, a naive real-time
QMC simulation scheme would have to perform a stochastic summation over 
all possible $\theta(x=0,t)$ paths. After discretization of the time line,
one is thus left with the task of sampling a high-dimensional integral
where each of the integration variables $\theta(0,t_j)$ 
corresponding to time slice $j$ is a $c$-number.

For numerical reasons, it is now advantageous to perform
the real-time simulation within an equivalent discrete model (a ``charge
representation'').  
Discrete variables significantly reduce the relevant configuration space 
subject to Monte Carlo sampling compared to the 
$\theta$ variables which are continuous. Their use constitutes
the first major step of our partial summation scheme. One 
can generally transform Luttinger liquid transport
problems involving elastic scattering by short-ranged impurities into
a discrete representation.  This is achieved by
expanding the impurity propagator\cite{schmid83}, 
\begin{eqnarray}\nonumber &&
\exp\left[ -i V_0 \int_0^t dt' \;\cos(2\sqrt{\pi}\theta(0,t'))\right] =
\\ && \label{imp} \sum_{j=0}^\infty 
\int_0^t dt_{j} \int_0^{t_j} dt_{j-1} \cdots
\int_0^{t_2} dt_1 \; (-i V_0/2)^j \\ && \times \sum_{\{q_j \}} \exp\left[
2\sqrt{\pi} i \int_0^t dt' \; q_j(t') \,\theta(0,t')\right] \;,\nonumber
\end{eqnarray}
where $q_j(t')=\sum_{i=1}^j \xi_i \delta(t'-t_i)$
is built from the discrete variables $\xi_i=\pm 1$.

The discrete ``charges'' $\xi_i$
can be understood as Hubbard-Stratonovich fields. 
Introduction of these auxiliary fields permits us to perform the 
now Gaussian integration over the Luttinger liquid bosons $\theta$.
Naturally, such mapping onto a discrete charge representation 
is not confined to Luttinger-liquid-type electron-electron
interactions and can be done for any Coulomb interaction potential.  
Furthermore, one can use similar charge representations
based on (\ref{imp}) for a multi-barrier problem, allowing for studies of
resonant tunneling or related phenomena.
In the following, we will focus on Eq.(\ref{lag1}),
which describes a Luttinger liquid impinging on a single barrier.

The nonlinear conductance $G(T,V)=\partial I/\partial V$ for transport of the
Luttinger liquid through a barrier can be readily computed
from the bosonized current operator, $J= (e/\sqrt{\pi})
\;\dot{\theta}(0)$. In fact, the conductance is 
\begin{equation}
G(T,V) =(e/\sqrt{\pi})\; (\partial/\partial V)  \lim_{t\rightarrow \infty} 
	\langle \dot{\theta}(0,t) \rangle \; , \label{cond}
\end{equation}
where the average is taken using (\ref{lag1}).
In practice, the expectation value $\langle\dot{\theta}(0,t)\rangle$
reaches a plateau value after some time $t^*$, and if the
real-time QMC scheme is able to reach this plateau, the
conductance can be computed directly from the plateau value.
Since $t^*$ grows very large at extremely low temperatures,
our method becomes increasingly costly near zero temperature.
However, as shown below, the dynamical sign problem can be 
effectively circumvented to allow for studies in the 
interesting low-temperature scaling region. 

Evaluating (\ref{cond}) by QMC requires construction of a discretized
path-integral expression for $\langle \dot{\theta}(0,t) \rangle$, where
$t$ is the maximum length of the time line considered in the 
simulation.
Eliminating all $\theta$ degrees of freedom away from the barrier,
$\langle \dot{\theta}(0,t) \rangle $ is expressed as a weighted sum 
over all sets of forward-backward real-time paths
$\{ \theta_f(0,t'), \theta_b(0,t') \}$ with $t'$ going from $t'=0$ to $t$.
In the next step, we switch to the charge representation using Eq.(\ref{imp}).
One can then integrate
out the $\theta(0)$ paths as well at the  price of introducing the auxiliary 
discrete charge paths $q_f(t')$ and $q_b(t')$. 
The resulting effective action contains nonlocal
couplings between $\{q_f,q_b\}$ due to the 
eliminated $\theta$ modes.

Next we express $q_f$ and $q_b$ in terms of their
symmetrized and antisymmetrized linear combinations. 
This constitutes the second major step with regard
to the partial summation scheme. 
The symmetrized component corresponds to a quasiclassical degree of freedom
since it describes propagation along the diagonal of the reduced
density matrix.  The antisymmetrized path, on the other
hand, describes quantum fluctuations corresponding to the off-diagonal
elements of the reduced density matrix.  

The basic idea of the partial summation scheme is to split off part
of the Monte Carlo problem to be evaluated non-stochastically, while 
leaving the remainder for the stochastic summation.  This can be
achieved by blocking states together in the course of the Metropolis
walk: instead of sampling states individually, one  
non-stochastically probes the local surroundings of a given state
to account for its degree of destructive interference\cite{makprl}. Such a 
blocking strategy avoids regions with highly destructive interference
which do not contribute to the path integral in the first place
but are responsible for the dynamical sign problem. 
In the present case, this strategy can be implemented
very efficiently by noting that one can sum over all symmetrized components
exactly. This is possible since there
are no quasiclassical self-interactions in the effective action.
In the end, only the antisymmetric paths are left for the Monte Carlo.

Any antisymmetric path can now be characterized by 
``kinks'' $y_i$ and the
respective kink times $t_i$. In the QMC calculation, we have to 
discretize time into $N$ sufficiently small slices of length $\delta t= t/N$, 
and no more than one kink is allowed on each time slice.
The possible kink 
values are $y_i=0,\pm 1,\pm 2$ corresponding to the 
charges $\xi=\pm 1$ of the original forward-backward charge paths. 
The Monte Carlo algorithm then performs a stochastic summation over 
all possible kink configurations $\{ y \}$. 
We have used both kink migration and double kink flip moves 
to generate the Metropolis trajectory. The MC weight function is defined 
such that $\langle \dot{\theta}(0,t')\rangle$ is 
evaluated only at the endpoint $t'=t$.  Our algorithm channels all
efforts into generating relevant configurations at this one time point alone.

To check our QMC simulation scheme, we have carried out two
independent tests: (a) short-time exact enumeration results are
accurately reproduced, (b) additional QMC simulations for the  static noise
$\langle I^2 \rangle$ show that  the familiar Johnson-Nyqvist formula 
is fulfilled at all temperatures. 
The results presented below 
employed discretizations on the order of $V_0 \delta t
\approx 0.3$, and an upper time limit of $V_0 t= 20$ was 
sufficient to reach the plateau region of $\langle \dot {\theta} (0,t)
\rangle$ for the temperatures studied here. 
Typically, numerical results for the low-temperature
region were obtained using 25 to 250 million passes and our
code performs at an average speed of 1 CPU hour per million passes
on an IBM RISC 6000 Model 590. 

Next we present numerical results for transport of a Luttinger liquid
through a barrier.  Figures \ref{fig1} and \ref{fig2} show simulation data
for the linear conductance $G$ at the two  
interaction strengths $g=1/3$ and $g=2/3$, respectively.
We have studied two different barrier heights, 
$V_0/\omega_{\rm c}=1/6$ and $V_0/\omega_{\rm c}=1/3$.
The value $g=1/3$ is of direct relevance to tunneling
of edge state excitations in the fractional quantum Hall 
effect\cite{fls95,moon93,mill}, while recent transport experiments in
quantum wires\cite{ths95} have been carried out at $g=0.67\pm 0.03$.  
Our results are of immediate interest for these experiments.

The main panels show the QMC data over a wide range of temperatures.
We compare our data to the thermodynamic Bethe ansatz solution
by Fendley, Ludwig, and Saleur (FLS)\cite{fls95} which
is exact in the weak-barrier case $V_0/\omega_{\rm c}\ll 1$.
The Kondo temperature is defined as\cite{fls95}
\[
k_{\rm B} T_{\rm B}/V_0 = t_g (\pi V_0/\omega_{\rm c})^{g/(1-g)} \;, 
\]
with $t_{1/3}=1.19599$ and $t_{2/3}=0.80549$.
The scaling variable used in the figures is
\[
X=d_g (T_{\rm B}/T)^{1-g}\;,
\]
where $d_g$ is readily determined from
the golden rule limit\cite{kf92} (specifically, 
$d_{1/3}=0.74313$ and $d_{2/3}=0.99806$).
The FLS solution reproduces the 
golden rule behavior 
\vbox{
\begin{figure}
\epsfig{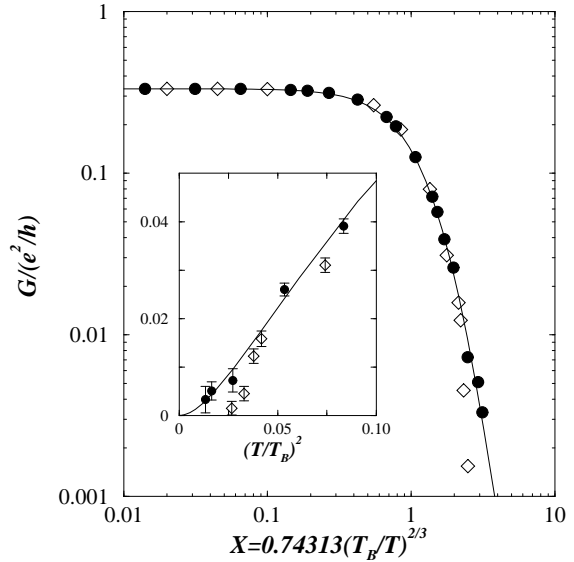}{3.00in}{0.00in}{30 100 555 635}
\caption[]
{\label{fig1}
QMC data for the linear conductance $G(X)$ at $g=1/3$.
Barrier heights are $V_0/ \omega_{\rm c}=1/6$ (circles)
 and $V_0/\omega_{\rm c}=1/3$ (diamonds). The solid curve is the
 FLS solution\cite{fls95}.
Inset: Low-temperature behavior. Error bars denote one standard
deviation, and the solid curve is again the FLS solution.
}
\end{figure}
\begin{figure}
\epsfig{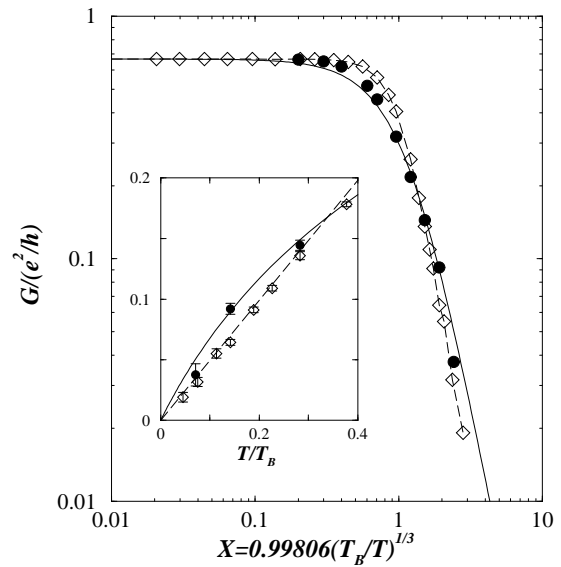}{3.00in}{0.00in}{30 100 555 635}
\caption[]
{\label{fig2}
QMC data for $G(X)$ at $g=2/3$.
Barrier heights are $V_0/\omega_{\rm c}=1/6$
(circles) and $V_0/\omega_{\rm c}=1/3$ (diamonds).
The solid curve is the FLS solution, the
dashed curve is a guide to the eye only.
Inset: Low-temperature behavior.
}
\end{figure}
}
$G/(e^2/h) = g (1- X^2)$ at $T\gg T_{\rm B}$ 
(small $X$), while for 
$T\ll T_{\rm B}$ it matches the KF scaling law (\ref{kfscale}),
\begin{equation} 
\label{keq}
G(X)/(e^2/h) = K_g X^{-2/g} \quad ({\rm large\,} X)\;,
\end{equation}
where $K_{1/3}=3.3546$ and $K_{2/3}=0.8177$.

For $g=1/3$, we compare our numerical results with the FLS curve 
at two values of the barrier height, $V_0/\omega_{\rm c} = 1/6$ and $1/3$
(see Fig.~\ref{fig1}).
For the smaller barrier, we reproduce the Bethe ansatz results well.  
This is quite remarkable since one might have expected more deviations 
from the FLS solution which assumes $V_0/\omega_{\rm c}\ll 1$.
In the FLS solution, the $T^4$ scaling region is found only at
very low temperatures $T/T_B$ below $\approx 0.15$.  This is corroborated
by our results which also indicate the {\em absence} of asymptotic 
$T^2$ contributions to the linear conductance.
The low-temperature KF scaling (\ref{kfscale}) thus seems to
hold even for the intermediate barrier heights studied here. 
Over the higher temperature range 
$0.15< T/T_{\rm B} < 0.4$, an {\em apparent}\, $T^2$
scaling is reached which, however, does not persist into the
asymptotic low-temperature regime.
In the case of the higher barrier, rather large deviations
from the FLS solution are seen, especially at low temperatures. 
Remarkably, the experimental data in Ref.\cite{mill} exhibit the same 
qualitative deviations.

In Fig.~\ref{fig2}, data are presented for $g=2/3$ corresponding to the
recent quantum wire experiments by Tarucha {\em et al.}\cite{ths95}.
Even for the intermediate barrier heights $V_0/\omega_{\rm c}=1/6$ and
$1/3$ considered here, we observe the 
$G\sim T$ scaling (\ref{kfscale}). While our data fully reproduce the 
FLS curve for $V_0/\omega_{\rm c}=1/6$, there are significant
deviations for a higher barrier. For $V_0/\omega_{\rm c}=1/3$, we find
a smaller value $K_{2/3}=0.477\pm 0.005$ instead of the FLS value $0.8177$,
but the low-temperature KF scaling (\ref{keq}) still holds.
The intermediate-barrier-height effects for $g=2/3$ and $g=1/3$
are qualitatively similar.

In conclusion, we have developed and applied real-time
quantum Monte Carlo simulations to low-temperature transport
of a Luttinger liquid through an arbitrarily high barrier.
The close agreement of our data with thermodynamic Bethe ansatz results
is rather remarkable since we have studied impurity levels
near the fermion band edge.  For $V_0/\omega_{\rm c}=1/6$, 
the FLS solution is completely reproduced, and only in the 
higher barrier $V_0/\omega_{\rm c}=1/3$ case are significant deviations
observed. 
Recent speculations about possible $T^2$ contributions to the 
linear conductance for intermediate barrier height 
are determined by our calculations to be incorrect, and 
the low-temperature $T^{2/g-2}$ scaling is supported for the full crossover
from a weak to an almost insulating barrier.
We believe that the method presented here will be useful in future studies
of nonequilibrium transport, noise properties, and resonant tunneling.

We wish to thank Paul Fendley for providing the Bethe ansatz curves
and other valuable inputs. We have benefitted from
discussions with H. Grabert, H. Saleur, M. Sassetti, and
U. Weiss.  This research is supported by 
NSF under grants CHE-9216221 and CHE-9257094, 
and by the Camille and Henry Dreyfus Foundation 
and the Alfred P. Sloan Foundation.  
Computational resources have been furnished by the IBM Corporation.

\end{document}